\documentclass{article}

\usepackage{PRIMEarxiv}

\usepackage[utf8]{inputenc} 
\usepackage[T1]{fontenc}    
\usepackage{hyperref}       
\usepackage{url}            
\usepackage{booktabs}       
\usepackage{amsfonts}       
\usepackage{nicefrac}       
\usepackage{microtype}      
\usepackage{lipsum}
\usepackage{fancyhdr}       
\usepackage{bm}
\usepackage{doi}
\usepackage{graphicx}       
\graphicspath{{media/}}     

\title{
A Pre-trained Deep Potential Model for Sulfide Solid Electrolytes with Broad Coverage and High Accuracy
}

\author{
  Ruoyu Wang\textsuperscript{1,2}\thanks{Those authors contribute equally to this work.}, Mingyu Guo\textsuperscript{3,4,*}, Yuxiang Gao\textsuperscript{1,2}, Xiaoxu Wang\textsuperscript{3,5}, Yuzhi Zhang\textsuperscript{3,5},\\ \textbf{Bin Deng}\textsuperscript{3}, \textbf{Xin Chen}\textsuperscript{6}, \textbf{Mengchao Shi}\textsuperscript{3,6}\thanks{shimengchao@dp.tech}, \textbf{Linfeng Zhang}\textsuperscript{3,5}\thanks{linfeng.zhang.zlf@gmail.com}   and \textbf{Zhicheng Zhong}\textsuperscript{1,2,6}\thanks{zczhong@ustc.edu.cn}\\
  \textsuperscript{1}School of Artificial Intelligence and Data Science, University of Science and Technology of China, Hefei 230026, China  \\
  \textsuperscript{2}Suzhou Institute for Advanced Research, University of Science and Technology of China, Suzhou 215123, China\\
  \textsuperscript{3}DP Technology, Beijing 10080, China  \\
  \textsuperscript{4}School of Chemistry, Sun Yat-sen University, Guangzhou 510006, China \\
  \textsuperscript{5}AI for Science Institute, Beijing 10080, China \\
  \textsuperscript{6}Suzhou Lab, Suzhou 215123, China}
 
\begin{document}
\maketitle
\begin{abstract}

Solid electrolytes with fast ion transport are one of the key challenges for solid state lithium metal batteries. To improve ion conductivity, chemical doping has been the most effective strategy, and atomistic simulation with machine-learning potential helps find optimized doping by predicting ion conductivity for arbitrary composition. Yet most existing machine-learning models are trained on narrow chemistry, and new model has to be trained for each system, wasting transferable knowledge and incurring significant cost. Here, we propose a pre-trained deep potential model purpose-built for sulfide electrolytes with attention mechanism, known as DPA-SSE. The training set encompasses 15 elements and consists of both equilibrium and extensive out-of-equilibrium configurations. DPA-SSE achieves a high energy resolution of less than 2 meV/atom for dynamical trajectories up to 1150 K, and reproduces experimental ion conductivity of sulfide electrolytes with remarkable accuracy. DPA-SSE exhibits good transferability, covering a range of complex electrolytes with mixes of cation and anion atoms. Highly efficient dynamical simulation with DPA-SSE can be realized by model distillation which generates a much faster model for given systems. DPA-SSE also serves as a platform for continuous learning, and the model fine-tune requires only a portion of downstream data. These results demonstrate the possibility of a new pathway for AI-driven development of solid electrolytes with exceptional performance.  

\end{abstract}

\section{Introduction}

Solid-state batteries are poised to revolutionize the energy storage industry due to their potential for higher energy density, excellent safety and improved sustainability, which are critical for addressing the limitations of current lithium-ion batteries \cite{janekSolidFutureBattery2016,janekChallengesSpeedingSolidstate2023}. Solid-state electrolytes is the key to the solid state batteries, and hence has attracted tremendous research interest\cite{zhaoDesigningSolidstateElectrolytes2020,famprikisFundamentalsInorganicSolidstate2019,liProgressPerspectiveCeramic2020,junLithiumSuperionicConductors2022,liLithiumSuperionicConductor2023,parkHighVoltageSuperionicHalide2020,wangProspectsHalidebasedAllsolidstate2022,tanCarbonfreeHighloadingSilicon2021}.  However, solid electrolytes still face critical challenges, particularly the low ion conductivity compared with liquid electrolytes. In this regard, sulfide electrolytes (e.g., LGPS-like and argyrodite) with fast ion transport are promising candidates for commercial application.\cite{katoHighpowerAllsolidstateBatteries2016,wenzelInterphaseFormationDegradation2016} Usually, the ion conductivity of base electrolyte can be further improved by chemical doping (i.e., element substitution), which has been one of the most effective strategies for electrolyte optimization towards high-energy storage.\cite{zhaoDesigningSolidstateElectrolytes2020,heHalogenChemistrySolid2023} With proper doping, an optimal lithium ion concentration with maximum ion conductivity can be achieved when there are both abundant diffusing ions and vacant hopping sites. High entropy materials with extensive doping also introduce local structure distortion, which disturbs the ion packing and removes the degeneracy of hopping barriers.\cite{zhaoDesigningSolidstateElectrolytes2020,zengHighentropyMechanismBoost2022} For instance, heavily doped sulfide electrolytes, such as the Li$_{9.54}$Si$_{1.74}$P$_{1.44}$S$_{11.7}$Cl$_{0.3}$\cite{katoHighpowerAllsolidstateBatteries2016}, the Li$_{9.54}$ [Si$_{0.6}$Ge$_{0.4}$]$_{1.74}$P$_{1.44}$S$_{11.1}$Br$_{0.3}$O$_{0.6}$\cite{liLithiumSuperionicConductor2023}, \textit{etc.}, achieve much higher ion conductivity than their base compounds. Therefore, material optimization with doping engineering is an important strategy to achieve the high-performance solid electrolytes.

Traditionally, wide chemical space has to be explored experimentally to find the optimized composition, which is both costly and time-consuming. Atomistic simulation of electrolyte materials with atomic details can significantly accelerate this process by predicting ion conductivity of arbitrary compositions\cite{vandervenRechargeableAlkaliIonBattery2020}. Atomistic simulation involves force field, which can be constructed either by fitting a physically informed model to experiments, or by first-principle quantum theory calculation. Empirical potentials, while being fast and efficient, are limited by their accuracy and transferability across different material configurations and temperature ranges. \textit{Ab initio} calculation achieves high accuracy, but the computational cost is exceedingly high. Thus the simulation is limited to a few hundred atoms and several nanoseconds in most cases, insufficient for practical simulation of solid electrolytes\cite{vandervenRechargeableAlkaliIonBattery2020}.

With the advent of machine learning (ML) potential like DeePMD\cite{zhangDeepPotentialMolecular2018,hanDeepPotentialGeneral2018}, \textit{etc.}, atomistic simulation with \textit{ab initio} accuracy but at the cost of empirical force field is possible\cite{behlerGeneralizedNeuralNetworkRepresentation2007,bartokGaussianApproximationPotentials2010,noeMachineLearningMolecular2020}. Trained over small scale \textit{ab initio} calculations, the computation cost of ML models only scales linearly with atom numbers, facilitating large scale simulation with high accuracy. Previous works have demonstrated the potential of ML models in solid electrolytes simulation\cite{leeDisorderdependentLiDiffusion2023}. Most importantly, ML-driven simulation reproduces ion conductivity of solid electrolytes with reasonable accuracy, which is previously intractable for either empirical or \textit{ab initio} force field\cite{huangDeepPotentialGeneration2021}. Nevertheless, the conventional ML force fields lack transferability over chemical and configuration space\cite{merchantScalingDeepLearning2023}. Consequently, new potentials have to be trained from scratch for each system, incurring significant cost in training and data generation. This issue is further exacerbated for solid electrolytes as it is necessary to simulate doped electrolytes with complex compositions. Universal force fields pre-trained across the period table, such as the DPA-2\cite{zhangDPA2UniversalLarge2023}, GNoME\cite{merchantScalingDeepLearning2023}, M3GNet\cite{chenUniversalGraphDeep2022}, CHGNet\cite{dengCHGNetPretrainedUniversal2023}, \textit{et al.}\cite{xieGPTFFHighaccuracyOutofthebox,yangMatterSimDeepLearning2024}, are highly transferable across vast composition space. However, the training sets of these universal models primarily consist of equilibrium configurations, and they generally struggle with out-of-equilibrium configurations. Unfortunately, ion transport calculation requires accurately predicting out-of-equilibrium configurations over dynamic trajectories, as a result, current universal models are unsatisfactory for solid electrolyte simulation. Given the limitation of current models, it is thus imperative to develop a new model scheme which is accurate while retaining the good transferability for solid electrolytes simulation.

Here, we introduce DPA-SSE, a pre-trained deep potential model with attention mechanism that is specifically designed for sulfide electrolytes simulation. The DPA-SSE model includes 41 unique systems, among which are 26 sulfide compounds, thereby providing a solid platform for the study of sulfide electrolytes and beyond. Trained over extensive out-of-equilibrium configurations, the model accurately predicts energy and atomic forces of sulfide electrolytes along dynamic trajectories, reproducing ion transport to \textit{ab initio} accuracy. The DPA-SSE exhibits good transferability. It accurately predicts solid solutions of arbitrary concentration, an important merit for sulfide electrolytes optimization. The pre-trained model can also be fine-tuned with minimal training data for downstream tasks. The fine-tuned model achieves similar performance with ten times less training data than models trained from scratch. Moreover, a distillation scheme that generates simpler models to alleviate the high cost associated with pre-trained universal models in dynamic simulation is proposed. Combining high precision and good transferability, the DPA-SSE could aid in the development of better sulfide electrolytes.

\begin{figure}
  \centering
    \includegraphics[width=0.8\textwidth]{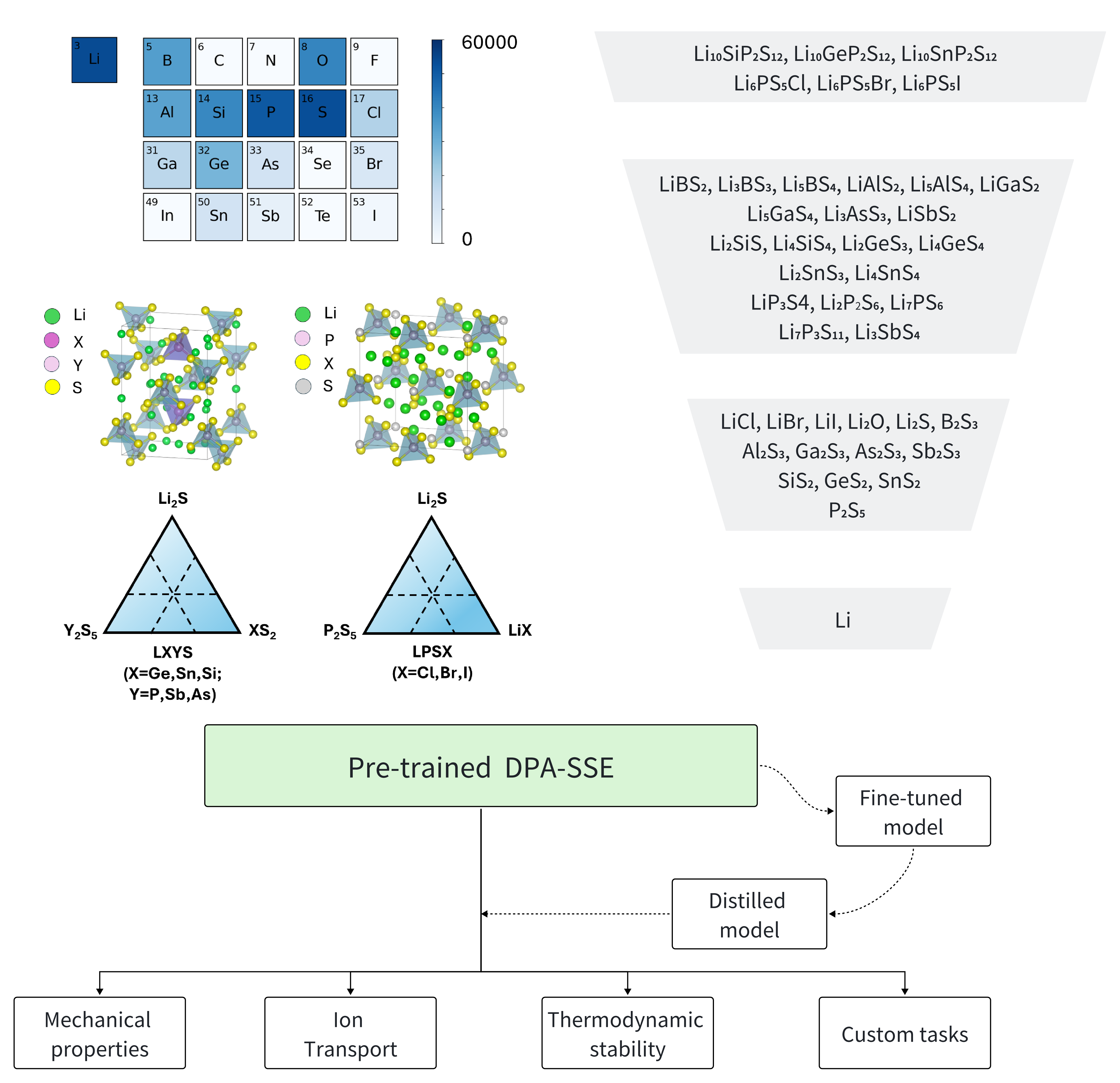} 
  \caption{\textbf{Model overview and training dataset of DPA-SSE model.} The DPA-SSE mainly covers LGPS-like and argyrodite electrolytes in the sulfide chemistry. The pre-trained model can be fine-tuned for downstream tasks with minimal additional data set. Through a distillation scheme, faster, lighter DeePMD models can be generated from the pre-trained or fine-tuned DPA-SSE for efficient molecular dynamics simulation.}
  \label{fig1}
\end{figure}

\section{Methods}
\subsection{Model design}
The DPA-SSE model utilizes the DPA-2\cite{zhangDPA2UniversalLarge2023} framework, a sophisticated model designed with broad coverage across the periodic table in mind. By leveraging the innovative architecture design, DPA-2 exhibits remarkable transferability with sufficient training sets. In DPA-2, a local descriptor $D_i$ is constructed from the local structural and chemical environment for each atom, represented by a single atom channel $f_i$, a rotation invariant atom pair channel $g_{ij}$ and a rotation equivariant channel $h_{ij}$. The single- and pair-atom representation subsequently passes through multiple transformer layers to output the local descriptor $D_i$ which is invariant under lattice translation, rotation and atom permutation. For each type of downstream task, a neural network would fit the local descriptor to the outputs. In the case of atomistic simulation, $D_i$ is fitted to a local energy $E_i$, whose summation over atomic indices gives the potential energy, $E(\bm{R})=\sum_i E_i(\bm{R})$, where $\bm{R}$ represents atomic configuration. The atomic forces are derived from the partial differentiation of potential energy with respect to atomic position, which naturally ensures force conservation.  

\subsection{Fine-tune and distillation}
The pre-trained DPA-SSE model can be fine-tuned for downstream tasks, wherein the atomic descriptors $D_i$ are initialized by the pre-trained model. Thanks to the transferable structure, the finetuned model converges much faster than models trained from scratch. Due to the large number of model parameters, direct application of the finetuned DPA-SSE to dynamic simulation for long time steps can be inefficient, which is a common issue for universal models. Thus, a \textit{distillation} strategy is devised. In this scheme, DPA-SSE model, the "teacher" model, essentially plays the role of an energy and force calculator, and a "student" model, usually a standard, lightweight DeePMD model without attention mechanism\cite{zhangDeepPotentialMolecular2018,zhangPretrainingAttentionbasedDeep2024,zhangDPA1PretrainingAttentionbased2023}, is trained with DPA-SSE as a labeler. A much faster model for dynamic simulation of certain compounds can be obtained. 

\subsection{Training data}
Figure \ref{fig1} provides a graphical representation of the training sets for DPA-SSE model. The training sets include three major types of sulfides electrolytes, namely, the LGPS-like compounds (Li$_{10}$XY$_2$S$_{12}$; X=Ge, Sn, Si; Y=P, Ga, Sb, As), argyrodites (Li$_{6}$PS$_5$X; X=Cl, Br, I) and Li$_7$P$_3$S$_{11}$. For solid batteries, the ability to operate at elevated temperatures is highly desirable, and MD simulations provide critical information regarding thermal stability of solid electrolytes. To this end, decomposition products of sulfide SEs are also required. By thermodynamic analysis, LGPS-like compounds decompose into Li$_2$S, Y$_2$S$_5$ and XS$_2$, whereas X=Ge, Sn, Si and Y=P, Ga, Sb, As; likewise, argyrodites decomposes to Li$_2$S, P$_2$S$_5$ and LiX (X=Cl, Br, I). These compounds, as well as the lithium metal, are included in the training sets. Overall, the training sets encompass 15 elements and 41 distinct systems, including 26 chalcogenides, as presented in Figure \ref{fig1}. The training sets are produced by the standard DP-GEN workflow\cite{zhangActiveLearningUniformly2019}. Initial configurations are constructed by sampling a short AIMD trajectory starting from a randomly perturbed structure. In subsequent training iterations, the training sets are collected through NPT simulations conducted over a temperature range of 0 to 1200 K and a pressure range of 0 to 2 GPa for extensive sampling of out-of-equilibrium configurations. Such a wide temperature and pressure range ensures extensive sampling of configurations relevant to MD simulation at heating temperature. The final training sets consist of 54771 snapshots of various configurations with DFT energy, force and virial as labels. The labeling of training sets are carried out on VASP packages\cite{kresseEfficiencyAbinitioTotal1996a} utilizing projector-augmented plane-wave (PAW) pseudopotentials with PBESol exchange-correlation functional\cite{perdewGeneralizedGradientApproximation1996a}. The cutoff energy was set to 600 eV for plane-waves, and the convergence condition was set to 0.01 eV/ Å for atomistic coordination relaxation. The Brillouin zone samplings were controlled by KSPACING in order to ensure the same k-point grid for all systems, and it is set to be 0.3.

\subsection{Model training}
The collected training sets are utilized to train the DPA-SSE model as mentioned before. The descriptor has two major components, a representation initializer (repinit) layer and 12 representation transformer (repformer) layers. The repinit layer consists of three hidden layers with 25, 50 and 100 neurons, respectively. The repformer layers serve to learn complex information within the latent space of data sets, enabling increasing model performance with the number of data frames. The learned descriptors are fitted to downstream tasks by a fitting network of three hidden layers each with 240 neurons. The transformer module generally implies a lower learning rate for training stability\cite{liuUnderstandingDifficultyTraining2023}. A low initial learning rate of 0.001 is chosen, which decays to \(3.51 \times 10^{-8}\) after 12,000,000 learning steps at an interval of 60,000 steps. We utilize the DPA-2 framework for single-task training. Specifically, we set the cutoff radius for the repinit layer to 9.0 Å. The input dimension for single-atom representations is fixed at 8. For the embedding connection, we employ a multilayer perceptron (MLP) consisting of three layers with 25, 50, and 100 neurons, respectively. Both types of layers are configured with four attention heads. The fitting network is composed of a three-layer MLP, each layer containing 240 neurons. Our training strategy starts with a learning rate of \(2 \times 10^{-4}\), undergoing exponential decay every 10,000 steps, eventually reducing to \(3 \times 10^{-8}\) at the 2,000,000 step. The prefactors for energy, force, and stress are adjusted alongside the learning rate. Specifically, the energy (virial) prefactor is scaled from 0.02 (0.2) to 1, while the force prefactor is adjusted from 1,000 to 1. The more detailed training parameters can be found in our previous  paper\cite{zhangDPA2UniversalLarge2023}, and we have made all the training sets and models available through the AIS-Square platform at \href{https://www.aissquare.com/datasets/detail?pageType=datasets&name=Solid_State_Electrolyte&id=217}{link}. We have created a notebook to help readers learn how to use open models for property calculation and model evaluation ($\mathrm{https://bohrium.dp.tech/notebooks/71679486918}$).

\subsection{Ionic conductivity}
The mean square displacement (MSD) of lithium ions after time period $t$ can be calculated as $\mathrm{MSD}(t)=\frac{1}{N}\sum_i^N|r_i(t)-r_i(0)|^2$, where $N$ is the number of lithium ions. Assuming Brownian motion, the diffusion coefficient $D$ of lithium ions in 3-dimensional space can be deduced as $D=\lim_{t\to\infty}\frac{\mathrm{MSD}(t)}{6t}$. The Nernst-Einstein relation relates ionic conductivity $\sigma$ at temperature $T$ to the lithium diffusion coefficient if the correlation between ions is negligible, $\sigma(T)=\frac{(ze)^2}{Vk_BT}D$, where $V$ the volume of simulation cell, $ze$ the electronic charge for each lithium ion, and $k_B$ is the Boltzmann coefficient\cite{france-lanordCorrelationsIonPairing2019}. Ion conduction is a thermally activated process, and the activation energy $E_a$ can be extrapolated by fitting the modified Arrhenius relation to the calculated ion conductivity at various temperatures, $\sigma(T)=\sigma_0T^{m}\mathrm{exp}(\frac{-E_a}{k_BT})$, with $m$ typically equal to -1. The pre-factor $\sigma_0$ is related to ion hopping entropy, distance, as well as attempt frequency in the simple case of uncorrelated ion hopping\cite{famprikisFundamentalsInorganicSolidstate2019}. In practice, $\sigma_0T^m$ can be approximated as constant over temperatures.

\section{Results}
\begin{figure}
  \centering
    \includegraphics[width=0.9\textwidth]{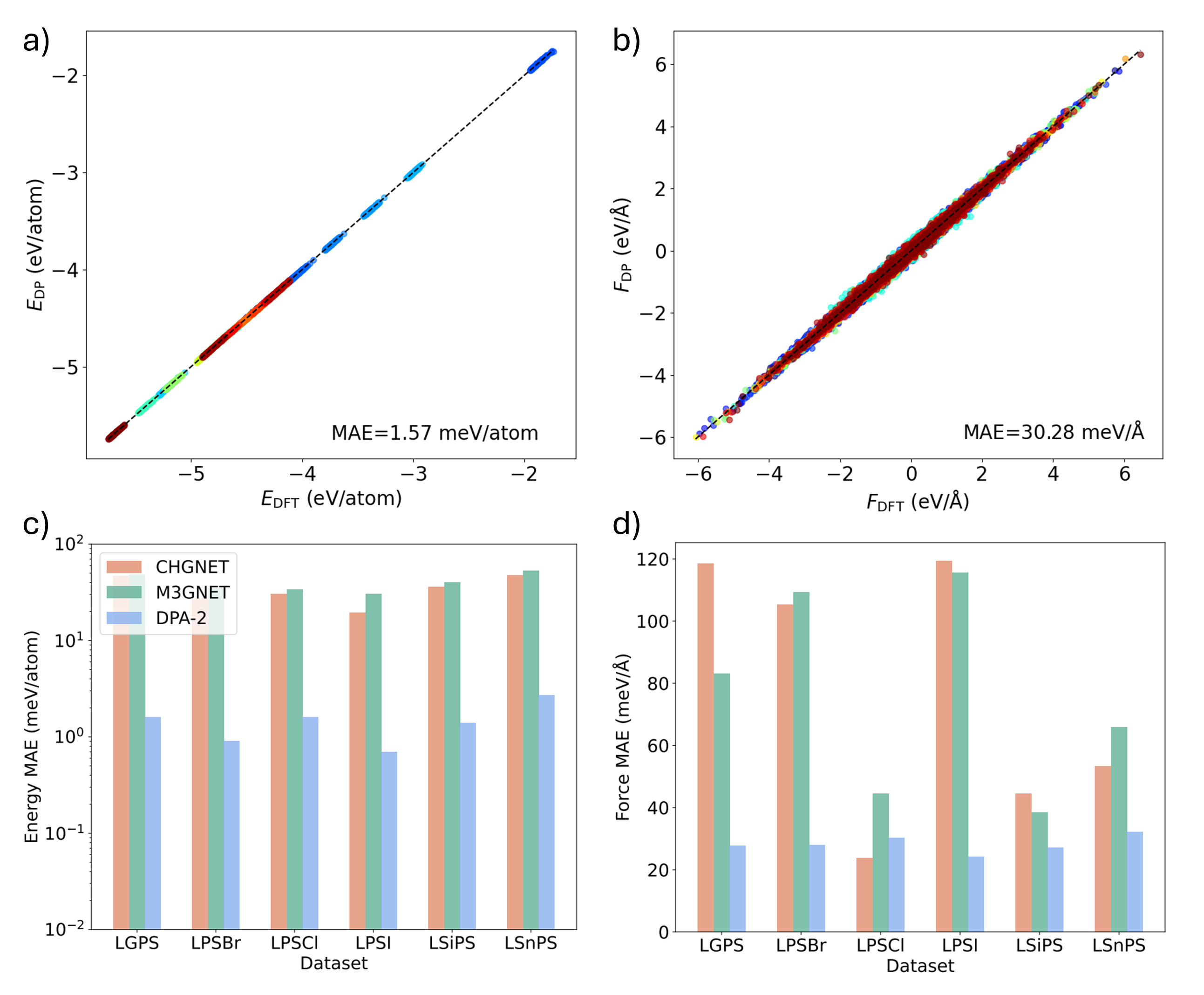}
 
  \caption{\textbf{Performance on test datasets.} (a-b) Mean absolute error (MAE) of energy and force near the equilibrium position compared to DFT calculations on the test datasets. Each color represents one of the 41 distinct training systems. (c-d) MAE of energy and force of DPA-2 and other universal IAPs on the heating DeePMD trajectories of typical sulfides electrolytes from 150 to 1150 K.}
  \label{fig2}
\end{figure}
\subsection{Model testing and performance }
DPA-SSE can accurately predict the energies and forces of sulfide electrolytes of various configurations. In Figure \ref{fig2}a and b, the predicted energy and force against DFT calculation are plotted. The energy and average force mean absolute error (MAE) converge to 1.58 meV/atom and 30.28 meV/Å on the test sets, respectively. For extensive model evaluation, additional tests are conducted to compare the DPA-SSE with other universal force fields, namely, the CHGNet and M3GNet. The test systems are typical sulfide electrolytes, Li$_{10}$XP$_2$S$_{12}$ (X=Ge,Sn,Si) and Li$_{6}$PS$_5$X (X=Cl,Br,I). Both M3GNet and CHGNet perform reasonably well near equilibrium positions, as shown in Figure \ref{test_equi}. However, DPA-SSE significantly outperforms other universal force fields on a long heating DeePMD trajectory from 150 to 1150 K, which includes extensive out-of-equilibrium configurations. The results are plotted in Figure \ref{fig2}c and d for energy and force prediction, respectively. A more detailed comparison of energy prediction by DPA-SSE and other universal force fields is presented in Figure \ref{test_900_diag}. The result suggests the energy is likely underestimated for out-of-equilibrium configurations by the universal force fields. Consequently, significantly higher ion transport may be obtained during dynamic simulations. The ability to accurately predict non-equilibrium configurations is absolutely crucial for the inference of key dynamic properties such as ion conductivity. The advantages of DPA-SSE in dynamic simulation can be partly attributed to the DP-GEN concurrent learning scheme which samples a considerable number of out-of-equilibrium configurations. DPA-SSE also accurately predicts lattice constants and formation energies of relevant systems, as shown in Figure \ref{latt}. The test result addresses the importance of training set construction for accurate simulation of solid electrolytes.

\begin{figure}
  \centering
    \includegraphics[width=1\textwidth]{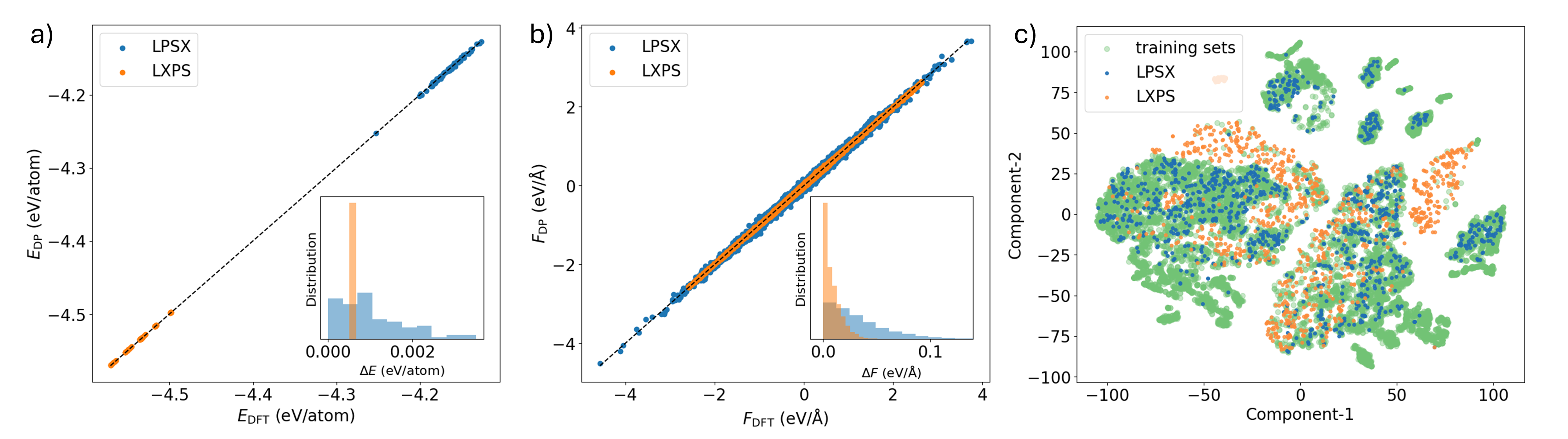} 
  \caption{\textbf{Transferability to solid solutions.} Mean absolute error and error distribution of (a) energy and (b) forces of Li$_{10}$Ge$_{1-x-y}$Sn$_x$Si$_y$P$_2$S$_{12}$ (LXPS) and Li$_{6}$PS$_5$Cl$_{1-x-y}$Br$_x$I$_y$ (LPSX) solid solutions. These compounds are not present in the training sets. (c) The t-distributed stochastic neighbor embedding (t-SNE) representation of the training set and the test sets for solid solutions. The t-SNE plot measures the similarity of two data points by their relative distance in a 2-dimension plane. According to the t-SNE plot, data points within the training set have covered main features of the solid solutions.}
  \label{fig3}
\end{figure}

\subsection{Model transferabiliity}
As mentioned in previous sections, chemical doping is a key strategy for electrolyte optimization, and atomistic simulations capable of predicting doped electrolytes may provide crucial guidance. Model transferability is essential here as no training set can cover the vast configuration space of solid solutions and doped compounds. To check the transferability of DPA-SSE, Li$_{10}$(Ge, Sn, Si)P$_2$S$_{12}$ (L(Ge, Sn, Si)PS) and Li$_{6}$PS$_5$(Cl, Br, I) (LPS(Cl, Br, I)) solid solutions with random mixes of cation and anion atoms at various concentration are constructed. For each configuration, an room temperature AIMD trajectory was produced and added to the test set. Figure \ref{fig3}a and b presents the energy and force prediction on the test sets. The energy and force MAE for solid solutions are 1.56 meV/atom and 29.15 meV/Å, respectively. The insets of Figure \ref{fig3}a and b also show the distribution of energy and force MAE. The prediction accuracy for solid solutions is on par with that for standard test sets as shown in Figure \ref{fig2}, suggesting good transferability over the solid solution case. To unveil the origin of predictability for solid solutions, the t-distributed stochastic neighbor embedding (t-SNE) plot for the data sets is presented in Figure \ref{fig3}c. The t-SNE plot projects high-dimension data distribution onto a 2D plane, visualizing the similarity of two data point by their distance. According to Figure \ref{fig3}c, points representing the solid solutions sit within the regions spanned by the training sets. Therefore, the transferability can be attributed to the efficient sampling strategy for training set and the advanced model architecture, which learned essential features of solid solutions from constituent compounds.
\begin{figure}
  \centering
    \includegraphics[width=1\textwidth]{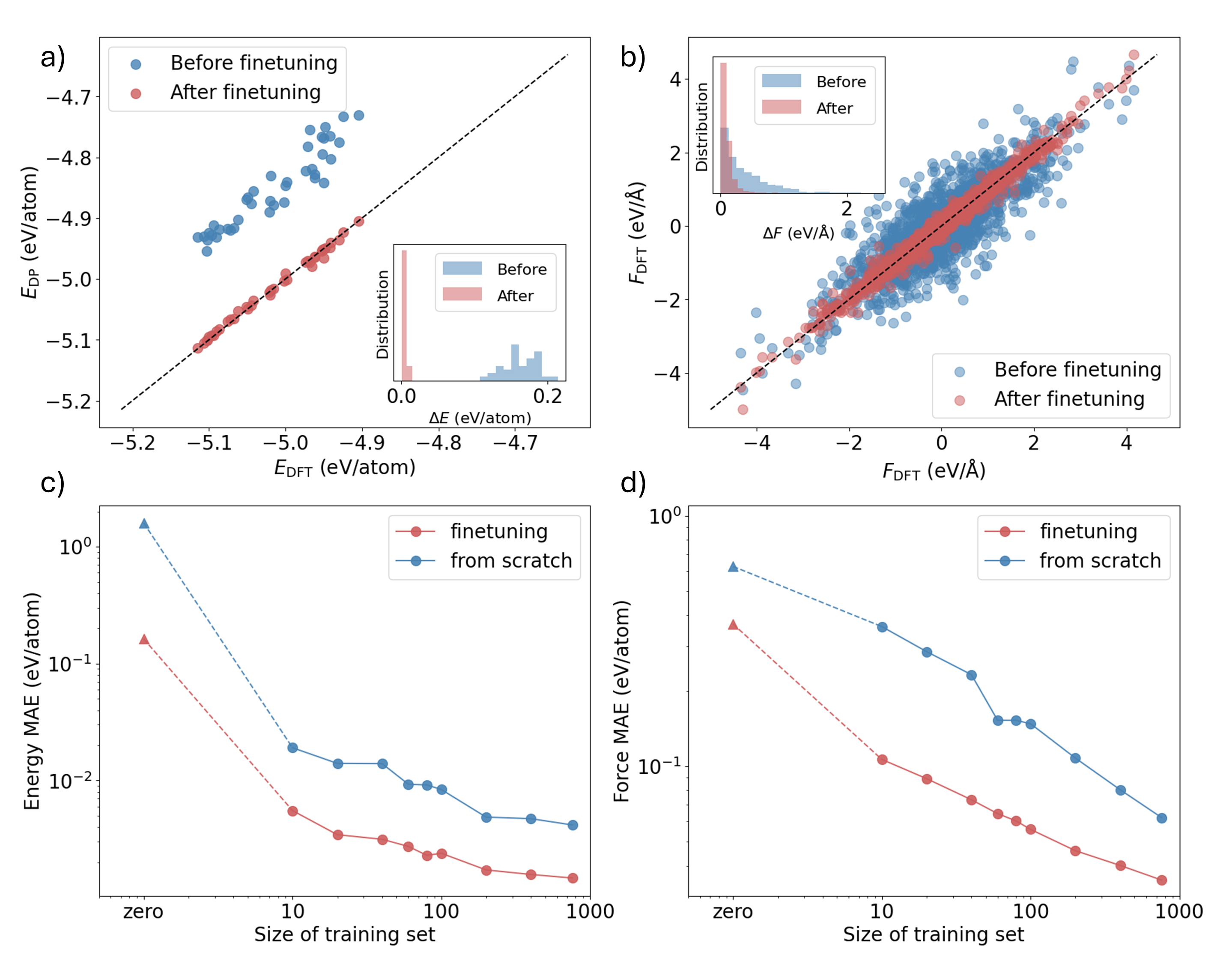} 
  \caption{\textbf{Efficiency of model finetuning.} The (a) energy and (b) force prediction of L$_2$B$_2$S$_5$ (mp-29410) systems before and after the model finetuning with 20 frames. The comparative sample efficiency on downstream task of finetuning and from-scratch training. (c) energy and (d) force prediction mean absolution error with the number of data frames.}
  \label{fig_ft}
\end{figure}

\subsection{Model fine-tune}
As well as accurate "zero-shot" prediction for materials with complex composition, a transferable model easily adapts to downstream tasks by model fine-tune, which requires much less training data than training from scratch. Here, we showcase the model fine-tune with DPA-SSE. 800 snapshots of a Li$_2$B$_2$S$_5$ (mp-29410, denoted as LBS) compound, which is not included in the training set (Figure \ref{LBS}), are collected from the DP-GEN workflow as downstream data set\cite{jansenNa2B2S5Li2B2S5Two1995}. The test set is constructed by 40 frames randomly chosen from the downstream data set. Figure \ref{fig_ft}a and b present the energy and force prediction on LBS before and after model finetuning. The "zero-shot" energy and force prediction MAE are 162.65 meV/atom and 369.16 meV/Å, respectively. An obvious bias can be observed for the zero-shot energy prediction. After fine-tuning with 20 randomly selected samples, the energy bias is eliminated, and the energy mean absolute error (MAE) is reduced to 3.44 meV/atom. The force prediction also significantly improves, achieving a MAE of 89.07 meV/Å. Figure \ref{fig_ft}c and \ref{fig_ft}d compare the energy and force convergence of randomly initialized and pre-trained models based on the size of training sets. Normalizing the training epochs by the number of data frames, we find that after fine-tuning with 60 frames, the pre-trained model outperforms the model trained from scratch using the complete training set of 760 frames. In this example, fine-tuning significantly reduces the amount of required training data, resulting in substantial savings on data generation. Model fine-tuning with DPA-SSE greatly accelerates the model training for unfamiliar compounds within the covered chemical space.

\subsection{Nudged elastic band calculation and dynamic simulations}
\begin{figure}
  \centering
    \includegraphics[width=1\textwidth]{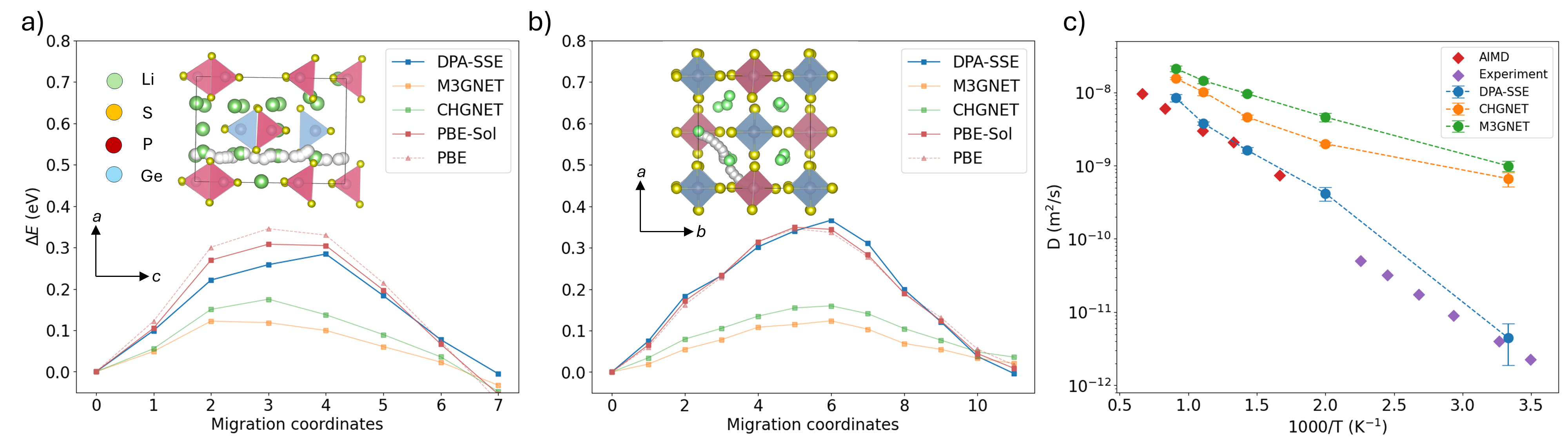} 
  \caption{\textbf{Nudged elastic band (NEB) calculation of concerted Li transport in Li$_{10}$GeP$_2$S$_{12}$ (LGPS) electrolytes.} Hopping barrier of (a) the concerted Li$^{+}$ ion motion along the \textit{c}-axis and (b) the Li$^{+}$ ion diffusion with a kick-off mechanism in the \textit{ab}-plane of LGPS electrolyte. The relative energies of frames along the hopping pathway are also predicted using DFT calculation as well as other universal force fields. Evidently, the predictions by DPA-SSE are more accurate than other force fields, which significantly underestimate the hopping barrier. (c) Experimental and calculated diffusion coefficient $D$ of LGPS electrolyte using AIMD, DPA-SSE and other universal force fields.}
  \label{fig5}
\end{figure}

Large-scale dynamic simulation of solid electrolytes for a sufficiently long time can predict ion transport properties. For precise ion transport calculations, it is essential that the underlying force fields can accurately evaluate out-of-equilibrium configurations during Li$^{+}$ ion hopping. Figure \ref{fig5}a and b present the hopping barrier for the correlated Li$^{+}$ motion along \textit{c}-axis and within the \textit{ab}-plane of LGPS electrolyte\cite{moFirstPrinciplesStudy2012,heOriginFastIon2017}. The diffusion path is optimized using nudged elastic band (NEB) method with the pre-trained DPA-SSE model. Along the NEB path, the Li$^{+}$ hopping energy is calculated as 0.28 eV along the \textit{c}-axis and 0.36 eV within the \textit{ab}-plane, which agrees well with the DFT result using PBESol functional. The other universal force fields, i.e., M3GNet and CHGNet, significantly underestimate the energy difference during Li$^+$ migration. As a result, the ion conductivity calculated by universal force fields may be significantly higher than the experimental values. To demonstrate the accuracy of DPA-SSE, Figure \ref{fig5}c compares the calculated and experimental lithium ion diffusion coefficient $D$ of LGPS from room temperature up to 1100 K. Clearly, diffusion coefficients calculated by DPA-SSE simulation are very close to that of AIMD simulation. Similar results are also obtained for other LGPS-like electrolytes, as shown in Figure \ref{LSiPS}. Noticeably, DPA-SSE enables accurate simulation of sulfide electrolytes at room temperature, which agrees reasonably well with available experimental results. This is previously intractable for AIMD simulation due to the long simulation time required for convergence and the large simulation cell to mitigate the size effect\cite{mortazaviMachinelearningInteratomicPotentials2020}. For comparison, general force fields overestimated the lithium ion diffusion, especially at ambient temperature. This result is consistent with the previous NEB calculation\cite{chenInsightIntrinsicInterfacial2017}.

\begin{figure}
  \centering
    \includegraphics[width=0.8\textwidth]{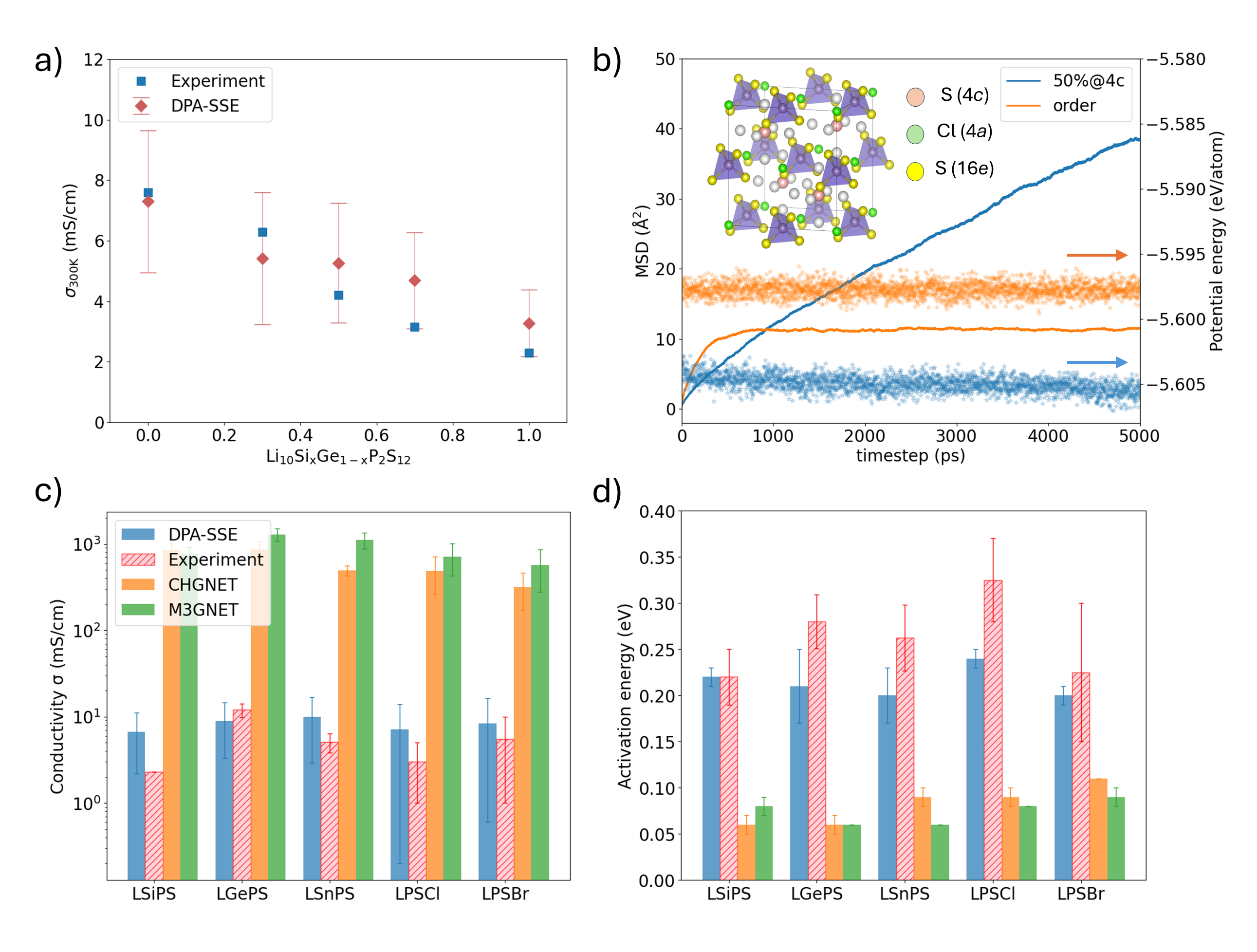} 
   \caption{\textbf{Molecular dynamic simulation of lithium ion transport in sulfide electrolytes.} (a) Experimental and calculated room temperature ion conductivity of Li$_{10}$Si$_x$Ge$_{1-x}$P$_2$S$_{12}$ solid solution at various concentrations. (b) The time evolution of lithium ion mean square displacement (MSD) for Li$_6$PSCl$_5$ in ordered and disordered anion-site configurations, the potential energy of the two configurations are also plotted. Experimental and calculated room temperature (c) ion conductivity and (d) hopping activation energy of five types of typical sulfide electrolytes.}
  \label{fig6}
\end{figure}

The ion transport of sulfide electrolytes can be modified by doping. Figure \ref{fig6}a plots the room temperature ion conductivity of Li$_{10}$(Si$_x$Ge$_{1-x}$)P$_{2}$S$_{12}$ solid solutions using DPA-SSE model. The result agrees reasonably well with experiment measurements. Even without external doping, there might be intrinsic disorder in the electrolyte materials which plays a key role in ion transport. The Li$^+$ ion transport in LPSX(X=Cl, Br, I) argyrodite electrolyte has been attributed to the S$^{2-}$/X$^{-}$ anion site disorder\cite{kraftInfluenceLatticePolarizability2017}. Powder X-ray diffraction analysis reveals that a portion of the halide atoms at the 4\textit{c} sites randomly exchange with an equal number of S atoms at the 4\textit{a} sites, as shown in the inset of Figure \ref{fig6}d\cite{morganMechanisticOriginSuperionic2021}. Molecular dynamic simulations with DPA-SSE capture the correlation between anion site disorder and ion transport\cite{leeDisorderdependentLiDiffusion2023}. Figure \ref{fig6}b plots the time evolution of the room temperature mean square displacement (MSD) of lithium ions in LPSCl with and without anion disorder. The lithium MSD converges after a few hundred picoseconds in the ordered phase; in the disordered phase, the lithium MSD increases with simulation steps, suggesting long-range ion transport. Hence, anion site disorder is essential for fast ion transport in argyrodites. Figure \ref{fig6}d also plots the potential energy of ordered and disordered LPSCl along simulation trajectory. The disorder phase actually has lower potential energy, and is thus more favorable than the ordered phase. This result is in agreement with experiment synthesis of LPSCl electrolytes. Contrary to LPSCl, the LPSI in ordered phase exhibits lower potential energy as shown in Figure \ref{LPSI}, which helps explain the significantly lower ion conductivity.

Considering the anion site disorder in argyrodites, Figure \ref{fig6}c compares room temperature ion conductivity of five sulfide electrolytes calculated with DPA-SSE, AIMD, and universal force fields. The results show excellent agreement with experiments, making significant improvement over the universal force fields trained on standard datasets\cite{chenUniversalGraphDeep2022,dengCHGNetPretrainedUniversal2023}. Assuming Arrhenius relations, activation energy $E_a$ of lithium ion diffusion can be extracted from the temperature dependence of diffusion coefficient $D$. The hopping activation energy is plotted in Figure \ref{fig6}d. Evidently, the hopping barrier $\Delta E$ estimated by DPA-SSE is very close to experimental measurement, whereas both universal force fields underestimate $\Delta E$. This result is consistent with the NEB calculations of Figure \ref{fig5}.

\subsection{Model distillation}
\begin{figure}[!htb]
  \centering
    \includegraphics[width=0.4\textwidth]{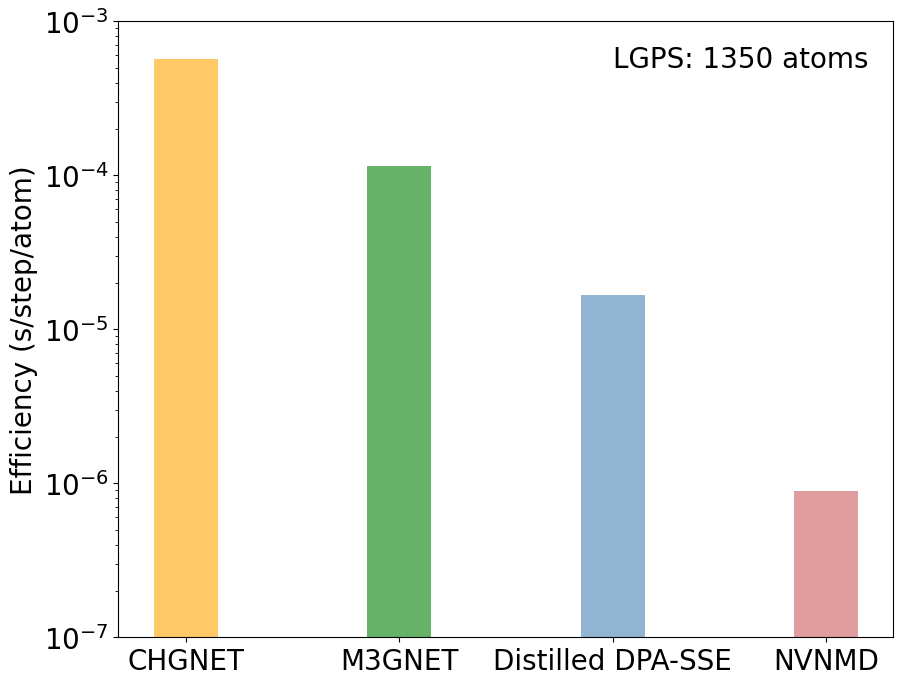} 
  \caption{\textbf{Model efficiency for molecular dynamic simulation.} The distilled DPA-SSE and both universal force fields are tested on a computation node with a single V100 GPU unit and 12 CPU cores. The NVNMD model is tested on custom hardware with a specialized architecture for running molecular dynamics.}
  \label{fig7}
\end{figure}

Despite the high accuracy and transferability, the DPA model is still rather expensive for dynamic simulations of long time steps. Usually, the simulation systems only represent a small portion of the chemical and configuration space covered by the pre-trained model. In these cases, the complex model structure associated with the transferability becomes redundant and can be trade-off for better efficiency. To address this situation, a model distillation scheme for DPA model was proposed, as demonstrated for the LGPS system in Figure \ref{distill}a. Firstly, thousands of randomly perturbed structures are generated. To ensure the robustness of the distilled model for dynamic simulations, we run a few short molecular dynamic simulation using the pre-trained model starting from perturbed structures. The resulting trajectory was further perturbed to generate more out-of-equilibrium configurations. These configurations are labeled by the pre-trained model, which effectively acts as a "teacher" model, and the ones with excessive atomic forces (in this case more than 10 eV/Å) are filtered out. Then a simple DeePMD model without attention layer, a "student" model, is trained for a million steps on the collected training set. Figure \ref{distill}b and c compares the performance of the "teacher" and "student" models on the test set constructed from a 900 K AIMD trajectory. The distilled "student" model exhibits comparable performance with the "teacher" model, whose energy and atomic force MAE are 2.10 meV/atom and 48.44 meV/Å, respectively. The inset of Figure \ref{distill}b compares the running efficiency of the pre-trained and distilled models. The benchmark test involves running a 900 K NVT simulation of an LGPS supercell consisting of 1350 atoms, on the same GPU processing unit. The distilled model is faster than the pre-trained model by at least two orders of magnitude. Figure \ref{fig7} compares the efficiency of distilled DPA-SSE and other universal force fields. The efficiency of the "distilled" model can be further improved by hardware-acceleration, \textit{e.g.}, the non von Neumann molecular dynamic (NVNMD) proposed by Mo et al\cite{moAccurateEfficientMolecular2022}. The NVNMD is revised from standard DeePMD model, and addresses the "memory bottleneck" issue by deploying purpose-built computers.  

\section{Discussion and Conclusion}
Ion conductivity is one of the most essential parameters of solid electrolytes. Atomistic simulation can be a valuable asset in searching for better electrolyte composition by screening candidate materials based on predicted ion conductivity. This may significantly accelerate the development process of solid electrolytes. To this end, the underlying force field of atomistic simulation must fulfill two requirements: firstly, it must be highly accurate even along out-of-equilibrium dynamic trajectories; secondly, it must be transferable to simulate doped electrolyte materials of complex composition. Traditional machined learning potential or universal force fields trained on near-equilibrium data sets either lacks transferability or does not have sufficient accuracy. As shown in our comparative tests, the universal force fields overestimate the room temperature ion conductivity of sulfide electrolytes by almost two orders of magnitude, thus fall short of the required accuracy.

In this work, we introduced the pre-trained DPA-SSE built for accurate simulation of sulfide electrolytes. Leveraging the extensive out-of-equilibrium training set and advanced model architecture, DPA-SSE model exhibits high accuracy while retaining good transferability among sulfide electrolytes. We demonstrate the capability of DPA-SSE by the Li$^+$ hopping barrier calculation using NEB method and accurately reproduce the experimental ion conductivity of sulfide electrolytes even with complex composition and intrinsic disorder such as the L(Si, Ge, Sn)PS and LPS(Cl, Br, I) solid solutions. As well as direct application, DPA-SSE also serves as a basis platform for more specialized tasks. The model can be fine-tuned for downstream tasks, achieving the same performance as the model trained from scratch but with ten times less training data, as demonstrated in the example of LBS system. When the chemical and composition space of the target electrolyte can be narrowed, the pre-trained or fine-tuned DPA-SSE model can generate a faster, lighter DeePMD model through the distillation scheme. The distilled model is much more efficient in dynamic simulation as it truncates model parameters irrelevant to the target system. 

Despite these advancements, further improvements can be achieved through several efforts. First, DPA-SSE at current stage only encompasses sulfide electrolytes, model coverage and transferability may be further enhanced by incorporating other systems. Second, an automated workflow for model fine-tune and distillation may greatly enhance the accessibility of DPA-SSE for field experts and industry users. This workflow should automatically validate the quality of fine-tuned models and the robustness of distilled models. Beyond predicting bulk electrolyte properties, interface models can be generated from DPA-SSE. In solid batteries, the interfacial composition and structure between electrolytes and electrode materials often significantly deviates from bulk materials, which presents major challenges in ion conduction and chemical stability\cite{xiaoUnderstandingInterfaceStability2019}. Machine learning force fields may play a pivotal role in understanding interfaces, but model training is difficult due to the complexity of interface and the requirement for high accuracy. With the fine-tune workflow, interface models built upon the pre-trained DPA-SSE may greatly improve training efficiency and accelerate the interface design process. In conclusion, DPA-SSE enables accurate large-scale atomistic simulation of sulfide electrolytes within a wide chemical and configuration space. Our model not only accelerates the optimization of sulfide electrolytes, but also serves as a platform model towards widespread applications of AI-driven techniques in solid electrolyte development. 

\section*{Acknowledgments}
This work was supported in part by the
National Science and Technology Major Project (Grants No. 2023ZD0120702)
, National Key R\&D Program of China (Grants No. 2021YFA0718900 and No. 2022YFA1403000), Key Research Program of Frontier Sciences of CAS (Grant No. ZDBS-LY-SLH008) and National Nature Science Foundation of China (Grants No. 12304049). We thank Bowen Deng and Dr. Peichen Zhong, the authors of CHGNet, for inspiring discussions. We also appreciate valuable advice by Dr. Qisheng Wu. The computational resource was supported by the Bohrium Cloud Platform at DP technology.

\bibliographystyle{unsrt}  
\bibliography{SSE_paper}  

\newpage
\setcounter{secnumdepth}{0} 
\setcounter{figure}{0}
\renewcommand\thefigure{S\arabic{figure}}
\section{Supporting information}

\begin{figure}[!htb]
  \centering
    \includegraphics[width=0.8\textwidth]{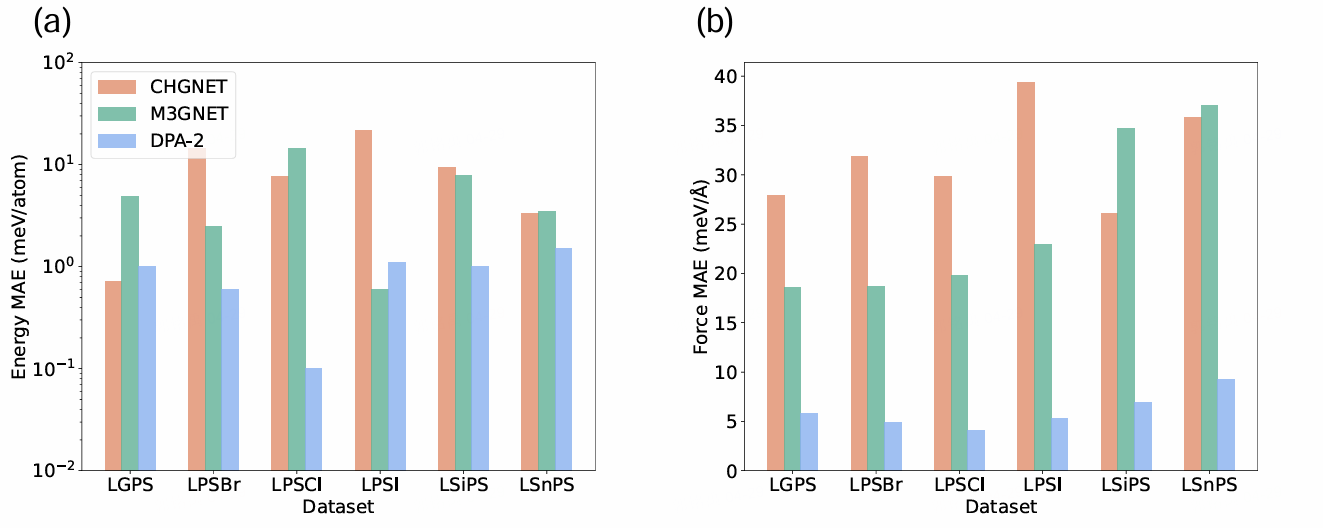} 
  \caption{The energy and force prediction error of DPA-2, M3GNet and CHGNet on near-equilibrium configurations of various sulfide electrolytes. The test sets are produced by running DeePMD simulation at 50 K for 10 steps.}
  \label{test_equi}
\end{figure}

\begin{figure}[!htb]
  \centering
    \includegraphics[width=1\textwidth]{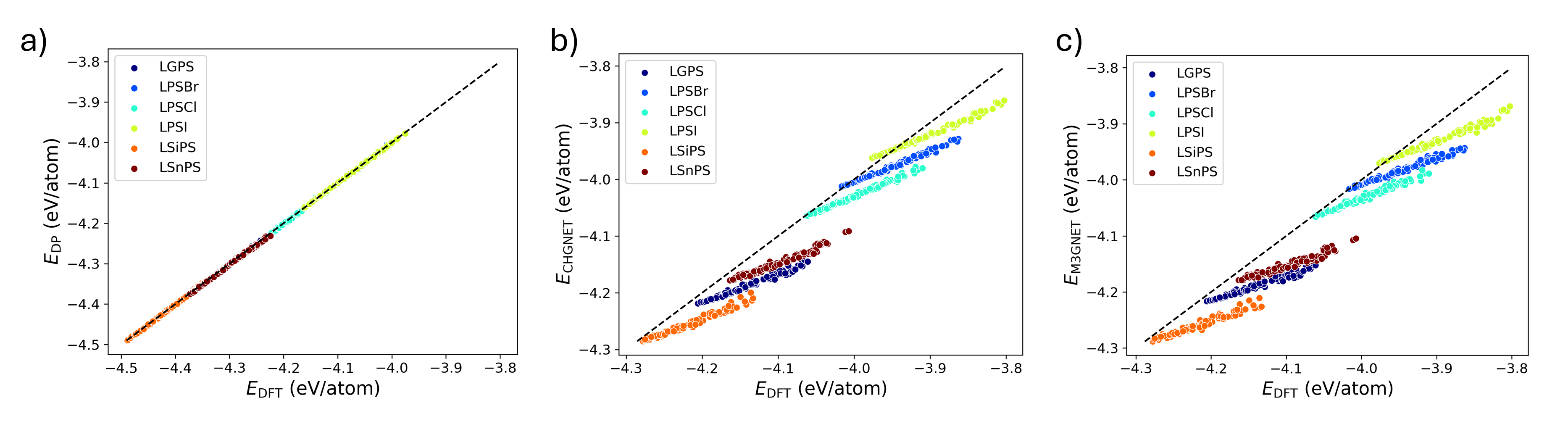} 
  \caption{The energy prediction error of DPA-SSE, CHGNet and M3GNet on a heating DeePMD trajectory from 150 to 1150 K.}
  \label{test_900_diag}
\end{figure}

\begin{figure}[!htb]
  \centering
    \includegraphics[width=0.75\textwidth]{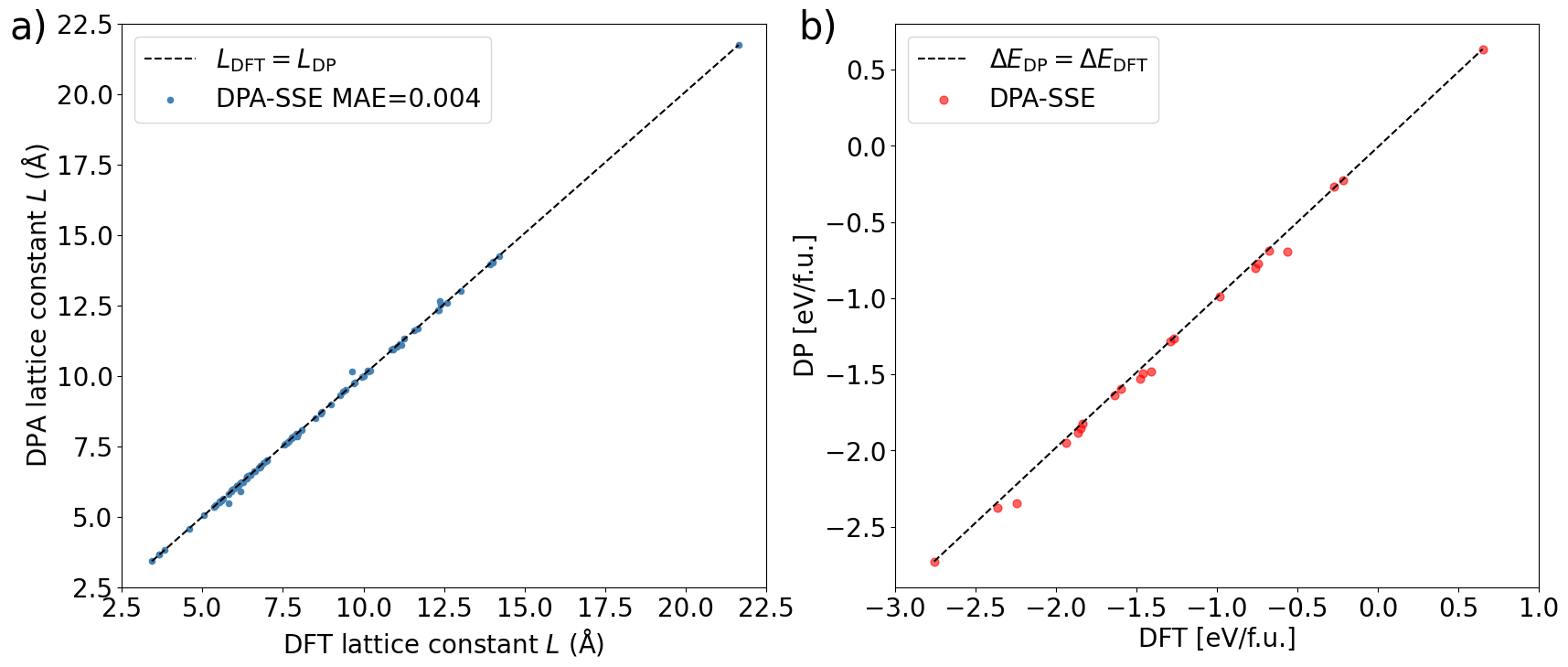} 
  \caption{Lattice constant and deformation energies. (a)The lattice constant of the 41 compounds covered by the training set as predicted by DPA-SSE and DFT calculation. (b)The decomposition energy of 22 compounds as predicted by DPA-SSE and DFT calculation}
  \label{latt}
\end{figure}

\begin{figure}[!htb]
  \centering
    \includegraphics[width=0.6\textwidth]{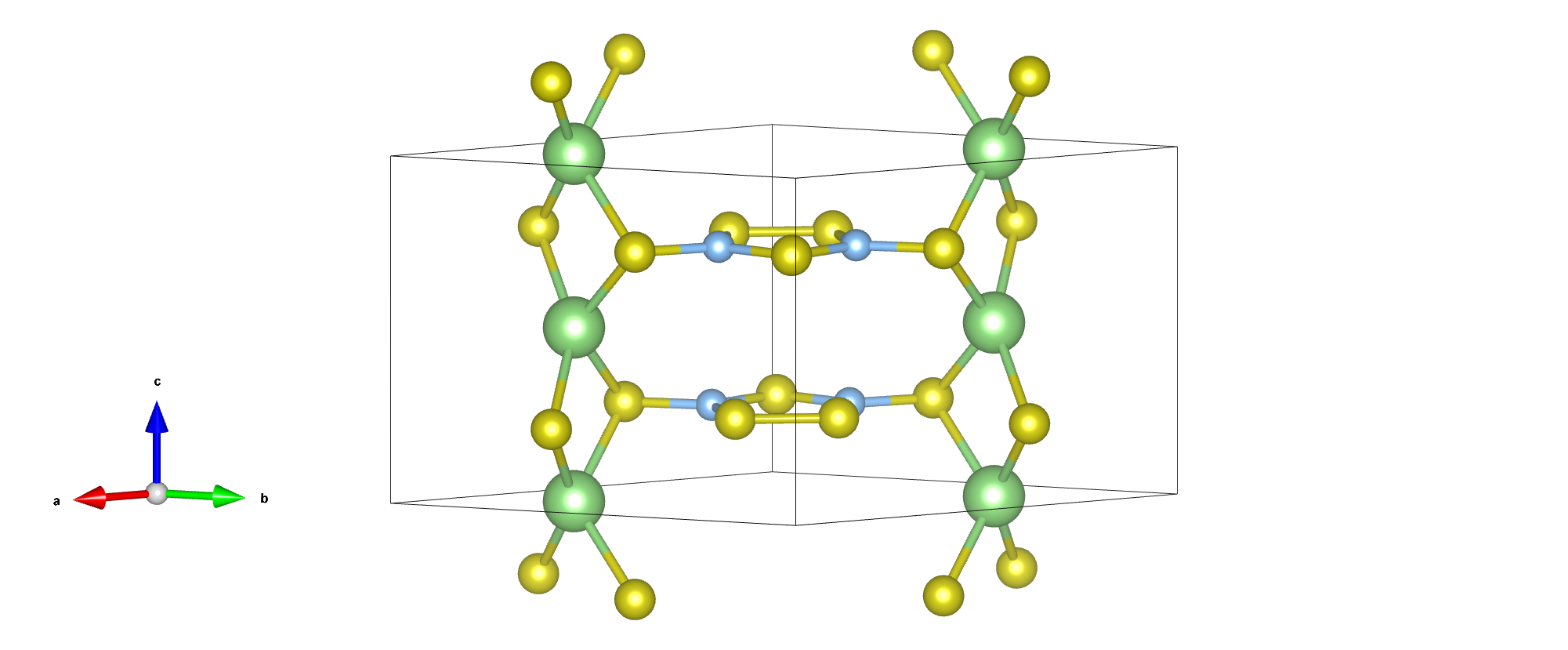} 
  \caption{The crystal structure of LBS mp-29410 compounds. The Li, B and S atoms are represented by green, blue and yellow colors, respectively.}
  \label{LBS}
\end{figure}

\begin{figure}[!htb]
  \centering
    \includegraphics[width=0.85\textwidth]{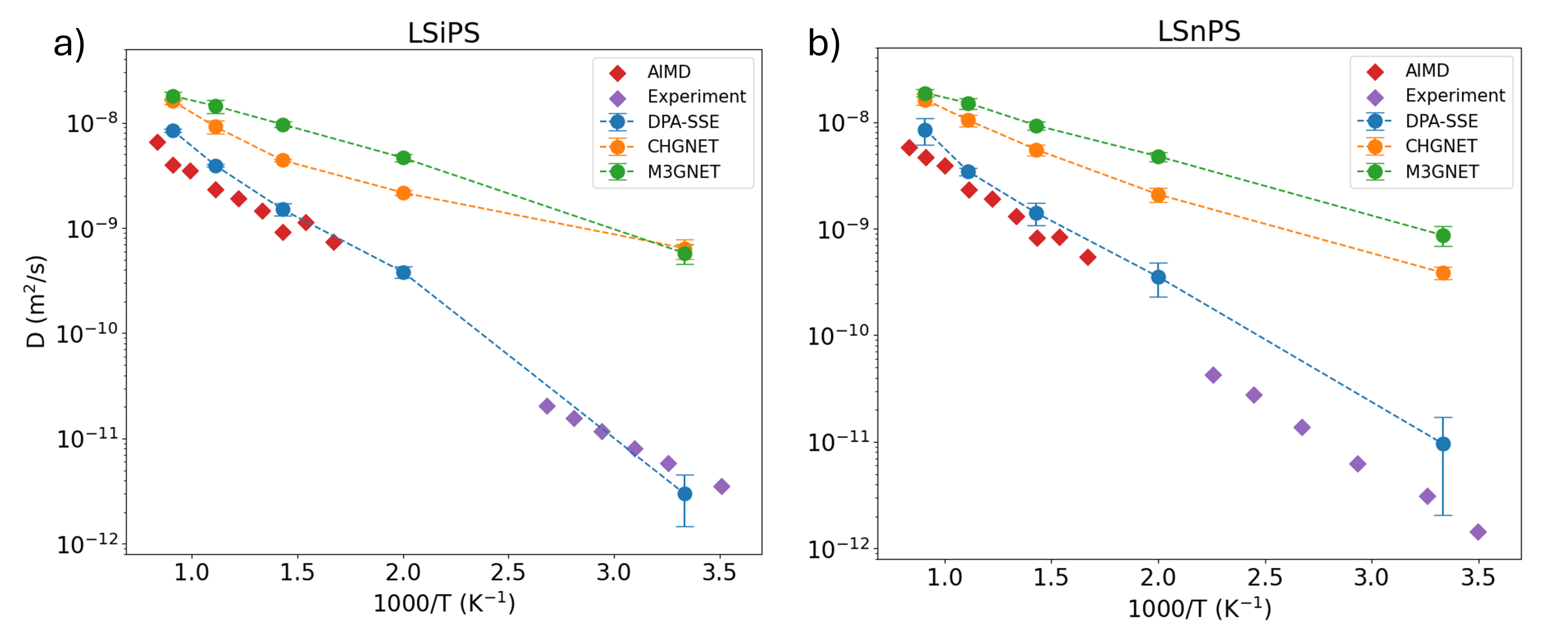} 
  \caption{The experimental and calculated diffusion coefficient of a) Li$_{10}$SiP$_2$S$_{12}$ and b) Li$_{10}$SnP$_2$S$_{12}$ solid electrolyte.}
  \label{LSiPS}
\end{figure}

\begin{figure}[!htb]
  \centering
    \includegraphics[width=0.5\textwidth]{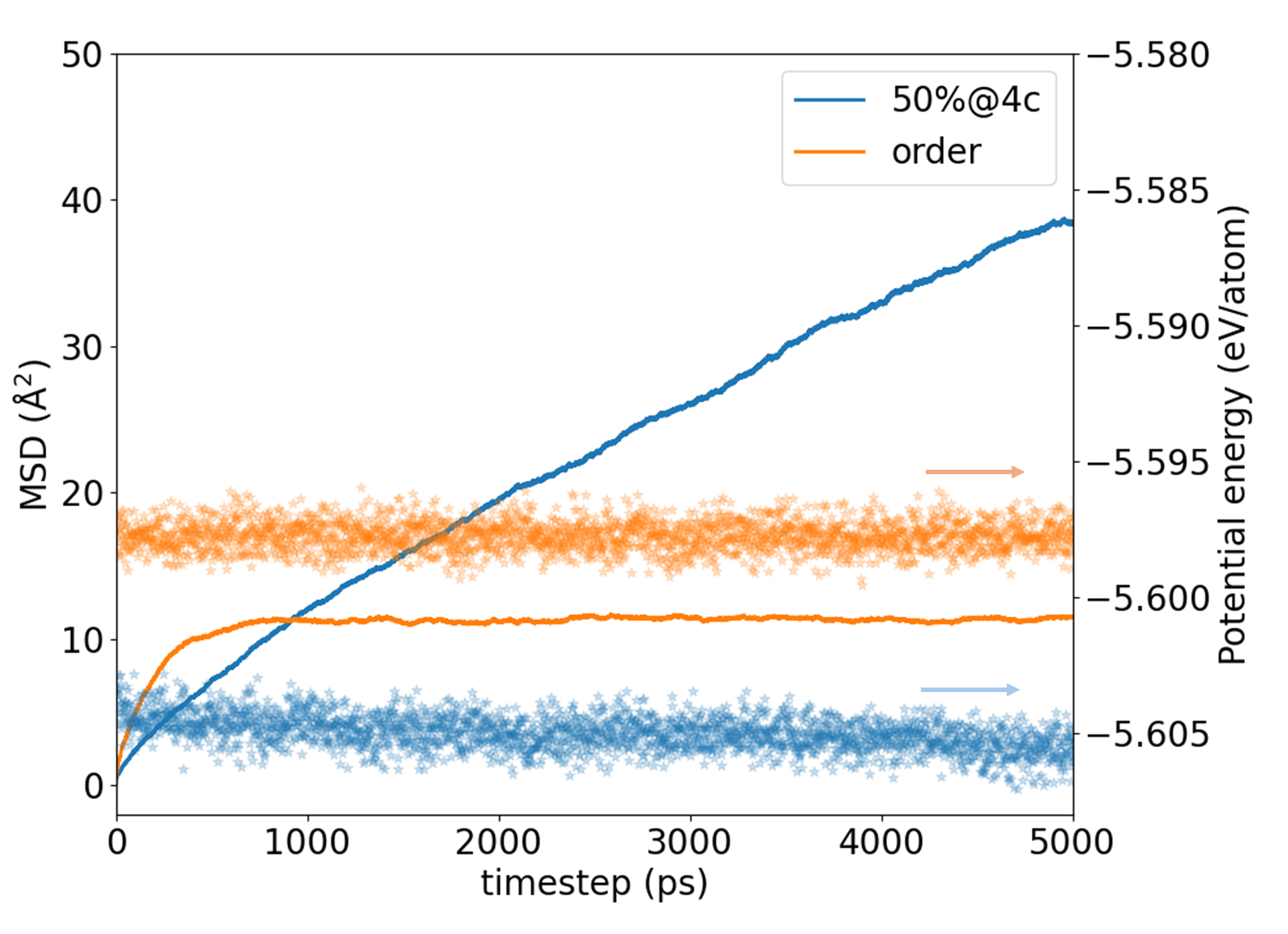} 
  \caption{The time evolution of room temperature mean square displacement (MSD) and potential energy of Li$_{6}$PS$_5$I solid electrolytes.}
  \label{LPSI}
\end{figure}

\clearpage
\newpage
\subsection{Model distillation}
The perturbed structures are first generated using the dpdata module, and then labeled by the pre-trained DPA-SSE model. Five perturbed structures are randomly selected as the initial structure for a short DPA-SSE molecular dynamics simulation of 2000 steps at 900 K using the NPT thermostat at ambient pressure. The time step of molecular dynamic simulation is 1 fs. One hundred frames are extracted from each trajectory, and each frame is further perturbed to generate more configurations. In total, the training set consists of 8579 frames of LGPS system. The test set is constructed from an 900K AIMD trajectory of NVT thermostat starting from the relaxed LGPS structure.

\begin{figure}[!htb]
  \centering
    \includegraphics[width=0.9\textwidth]{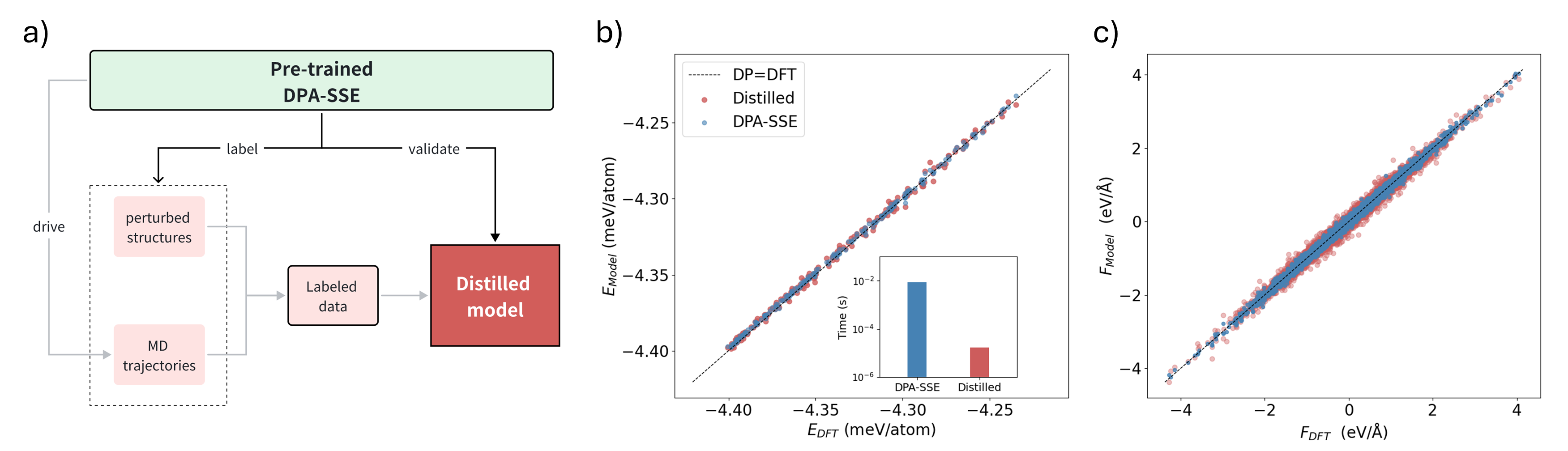} 
  \caption{\textbf{Model distillation for efficient dynamic simulation.} a) The schematic of the distillation process. The b) energy and c) force prediction for the pre-trained and distilled model on an AIMD trajectory of LGPS at 900 K. The inset of b) compares the efficiency of two models.}
  \label{distill}
\end{figure}

\end{document}